\documentclass[twocolumn,aps,showpacs,amsmath,amssymb,floatfix]{revtex4}
\usepackage{graphicx}
\usepackage{array}
\usepackage{float}
\begin{document}
\title{Winning quick and dirty: the greedy random walk}
\author{E.~Ben-Naim}\email{ebn@lanl.gov}
\author{S.~Redner}\email{redner@lanl.gov}\altaffiliation{Permanent address: 
Department of Physics, Boston University, Boston, Massachusetts, 02215 USA}
\affiliation{Theory Division 
and Center for Nonlinear Studies, Los Alamos National Laboratory, Los
Alamos, New Mexico 87545 USA}
\begin{abstract}
  As a strategy to complete games quickly, we investigate
  one-dimensional random walks where the step length increases
  deterministically upon each return to the origin.  When the step
  length after the $k^{\rm th}$ return equals $k$, the displacement of
  the walk $x$ grows linearly in time.  Asymptotically, the
  probability distribution of displacements is a purely exponentially
  decaying function of $|x|/t$.  The probability $E(t,L)$ for the walk
  to escape a bounded domain of size $L$ at time $t$ decays
  algebraically in the long time limit, $E(t,L)\sim L/t^{2}$.
  Consequently, the mean escape time $\langle t\rangle\sim L\ln L$,
  while $\langle t^n\rangle\sim L^{2n-1}$ for $n>1$.  Corresponding
  results are derived when the step length after the $k^{\rm th}$
  return scales as $k^\alpha$ for $\alpha>0$.

\end{abstract}


\maketitle

\section{Introduction}

A popular card game for young children is ``war''.  The rules of this
game are extremely simple: with two players, start by dividing the
cards evenly between the players.  At each play, the players expose
the top card in their piles.  The person with the higher card collects
the two exposed cards and incorporates them into his or her pile.
When there is a tie, a ``war'' ensues. Each player contributes three
additional cards from his or her pile to the pot and then the fourth
card is exposed.  The winner takes the pot. In case of another tie,
the war is repeated until there is a winner.  The game ends when one
player no longer has any cards.

Similar to many other games including for example, coin toss,
roulette, and dreidel, the game of war resembles a random walk
\cite{HG,LPP,DD}, as the number of cards possessed by each player
changes by $\pm 1$ (or by $\pm 5$, $\pm 9$, {\it etc.}, when
occasional wars occur) after each play.  Since there are $N=52$ cards
in a deck, a natural anticipation is that the length of the game
scales as $N^2$ \cite{ruin,RV}.  Based on soporific experiences in
playing war with our children, it is desirable to modify the rules so
that the game ends more quickly.  We have found that the following
modification -- which we term ``superwar'' -- works quite well: in
each war, increase the number of cards that a player contributes to
the pot by one compared to the previous war.  This modification of war
ends much more quickly than the original game and is also more
exciting for young children.

The game of superwar inspires the present work in which we investigate
the properties of a one-dimensional random walk in which the step
length increases in a deterministic manner each time the walk returns
to the origin.  We term this process the Greedy Random Walk. More
generally, we consider the situation where the step length $\ell_k$
after the $k^{\rm th}$ return to the origin is
\begin{equation}
\label{ell}
\ell_k=k^\alpha,
\end{equation}
with positive $\alpha$; the initial step length equals one.  The increasing
step length corresponds to increasing the payoff when the cumulative score of
a game is tied.  This mechanism provides a strategy to complete games
quickly, although it differs than superwar where the payoff is raised when
there is a tie in a single play.

In the next section, we give heuristic arguments for the typical displacement
and for extremal properties of the probability distribution.  Then we study
the probability distribution of the greedy walk in an infinite system.  We
present simulation results as well as an asymptotic solution for this
distribution.  Our solution relies heavily on classic first-passage
properties of random walks \cite{F68,W,fpp,RG}.  The probability distribution
of the greedy random walk has several intriguing non-scaling features,
including sharp valleys at the prime numbers and an anomalous contribution
due to walks that never return to the origin.

In Sec.~III, we determine how long it take for a greedy walk to escape the
finite interval $[-L,L]$.  Generically, the escape time grows more slowly
than the diffusive time scale $L^2/D$.  As a consequence, the game ends much
more quickly.  However, the escape probability has a power-law tail so that
the higher moments are controlled by the diffusive time scale.  Thus the
escape time of the greedy walk are characterized by large fluctuations.

\section{Displacement statistics}

\subsection{Typical and Extremal Displacements}

We first determine the typical displacement of the greedy random walk.
A crucial fact is that the statistics of returns to the origin are not
affected by the growth of the step length.  Thus, a typical walk of
$t$ steps will visit the origin of the order of $k\sim t^{1/2}$ times
\cite{F68,W,fpp,RG}.  Thus, the step length $\ell=k^\alpha$ grows as 
$t^{\alpha/2}$.  For an ordinary random walk with step length $\ell$,
the typical displacement is $x\sim t^{1/2}\ell$.  Combining these two
scaling laws, the typical displacement of the greedy walk grows with
time as
\begin{equation}
\label{typical}
x\sim t^\nu,
\end{equation}
with $\nu={(1+\alpha)/2}$.  Thus the greedy walk is more extended than
a conventional random walk.

We expect that this typical displacement characterizes the probability
distribution that the greedy walk is at position $x$ at time $t$,
$G(x,t)$.  In the long time limit, this distribution should thus
conform to the conventional scaling form
\begin{eqnarray}
\label{fz}
G(x,t)\simeq t^{-\nu}\Phi\left(x\,t^{-\nu}\right),
\end{eqnarray}
where $\Phi(z)$ is the scaling function.

We can determine the asymptotic decay of $\Phi(z)$ by using a Lifshitz
tail argument \cite{MEF,L}.  As a preliminary, we need to identify the
walks with the maximal possible displacement.  For conventional random
walks, the extremal walk is ballistic -- stepping in one direction
only.  For the greedy random walk, in contrast, extremal walks involve
a compromise between returning to the origin often, so as to acquire a
large single-step length, and moving in one direction, so as to be as
far from the origin as possible.  We are thus led to consider a hybrid
zig-zag/ballistic walk that makes immediate reversals for the first
$\tau$ steps and then moves in one direction for the remaining
$t-\tau$ steps.

\begin{figure}[ht]
  \vspace*{0.cm}
  \includegraphics*[width=0.375\textwidth]{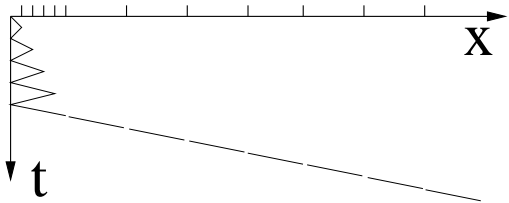}
 \caption{Extremal greedy walk of 16 steps when the step length grows
   linearly at each return to the origin.  There are 4 immediate reversals in
   the first 8 steps and then the remaining steps are all in one direction. }
 \label{opt}
\end{figure}

When only immediate reversals in direction occur, the step length
after $\tau$ steps ($\tau/2$ returns) is
$\ell_{\tau/2}=(\tau/2)^\alpha$.  Then if the remaining steps are all
in one direction, the displacement of the walk at time $t$ is
\begin{eqnarray}
\label{xmax}
x =  (t-\tau)\, \left(\frac{\tau}{2}\right)^\alpha.
\end{eqnarray}
Maximizing this expression with respect to $\tau$, the optimum value
of $\tau$ is $\tau=\frac{\alpha}{1+\alpha}\, t$ and the maximal
displacement is
\begin{eqnarray}
x_{\rm max} \propto t^{1+\alpha}.
\end{eqnarray}
Notice that the exponent characterizing the maximal displacement is
twice that of the typical displacement, $x_{\rm max}\sim t^{2\nu}$.

We now exploit the result for $x_{\rm max}$ to estimate the tail of
the probability distribution.  We first make the standard assumption
that the tail of the probability distribution decays as a stretched
exponential, \hbox{$\Phi(z)\sim \exp\left(-|z|^\delta\right)$}, for
$z\gg 1$ \cite{MEF}, where all factors of order one have been ignored.
With this ansatz, the probability for the maximal-displacement greedy
walk asymptotically scales as
\begin{eqnarray}
\label{P-ext}
G(x=t^{2\nu},t)\sim  \exp\left(-t^{\nu\delta}\right).
\end{eqnarray}
On the other hand, the probability for this maximal-displacement walk
decays exponentially with time, since a finite fraction of the steps
in the walk must be uniquely specified.  Equating this exponential
decay to the form given in Eq.~(\ref{P-ext}), we immediately conclude
that $\delta =1/\nu$.  As a result, we deduce that the scaling
function in $G(x,t)$ decays according to
\begin{eqnarray}
\label{fz-tail}
\Phi(z)\sim \exp\left(-|z|^{2/(1+\alpha)}\right),
\end{eqnarray}
for $z\gg 1$.  Notice that the conventional Fisher scaling relation
$\delta=(1-\nu)^{-1}$ is violated for greedy walks \cite{MEF}.

\subsection{The Probability Distribution}

\begin{figure}[ht]
  \vspace*{0.cm}
  \includegraphics*[width=0.4\textwidth]{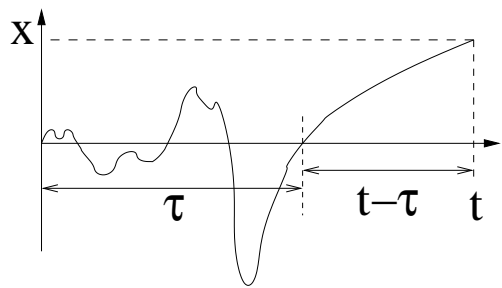}
 \caption{Space-time trajectory of an example greedy random walk, with 4
   returns to the origin after $\tau$ steps.  Upon each return, the step
   length increases, as indicated schematically.  After the last return, the
   final position is reached in the remaining $t-\tau$ steps.}
 \label{traj}
\end{figure}

We can obtain the full probability distribution of greedy walks by utilizing
basic first-passage properties of ordinary random-walks.  These first-passage
techniques provide an insightful and pleasant way to understand greedy walks.
There are two generic ways that the greedy walk can be at position $x$ at
time $t$ when starting at the origin at $t=0$.  The first is to reach $x$
without ever returning to the origin.  The second possibility, as depicted in
Fig.~\ref{traj}, is that the walk returns to the origin $k$ times, with the
$k^{\rm th}$ return occurring at time $\tau$, and then the walk reaches $x$
in the remaining $t-\tau$ steps without touching the origin again.  The
number of returns to the origin is variable, but the maximum number cannot
exceed $t/2$.

According to this decomposition of a greedy walk into a segment that consists
of $k$ returns to the origin and a non-return segment, we can write the
probability distribution of greedy walks in the following form:
\begin{equation}
\label{pxt-def}
G(x,t)= Q(x,t)+
\sum_{k=1}^{k_{\rm max}}\int_0^t\! F^{(k)}(0,\tau) \,\frac{1}{k^{\alpha}}
Q\!\left(\frac{x}{k^\alpha},t\!-\!\tau\right) d\tau.
\end{equation}
Here $k_{\rm max}$ is the maximum possible number of returns to the
origin in time $t$ when the final displacement is $x$.  The first term
on the right is the probability that a walk, which never returns to
the origin, is at $x$ at time $t$.  In the second term,
$F^{(k)}(0,\tau)$ is the $k^{\rm th}$-passage probability to the
origin, namely, the probability that the walk returns to the origin
$k$ times, with the $k^{\rm th}$ return occurring at time $\tau$.  The
term $\frac{1}{k^\alpha} Q\left(\frac{x}{k^\alpha},t-\tau\right)$ then
accounts for the probability for the walk to reach $x$ in the
remaining $t-\tau$ steps without touching the origin again.  Because
the step length is $k^\alpha$ in this last leg of the walk,
$x/k^\alpha$ steps are required to reach $x$.  We also need to include
the prefactor $1/k^{\alpha}$ to ensure proper normalization.

Since Eq.~(\ref{pxt-def}) is in the form of a convolution, it is much
more convenient to work with Laplace transforms.  Then the basic
equation for the probability distribution simplifies to
\begin{equation}
\label{pxs-def}
G(x,s)= Q(x,s)+ \sum_{k=1}^{k_{\rm max}} F^{(k)}(0,s) \,\,
\frac{1}{k^\alpha}Q\left(\frac{x}{k^\alpha},s\right),
\end{equation}
where the argument $s$ is generally used to signify Laplace
transformed quantities.  Each of the terms on the right-hand side of
this equation are well known first-passage properties
\cite{F68,W,fpp,RG}, from which we can then obtain the probability
distribution of greedy walks.

We now determine the individual terms that appear in
Eq.~(\ref{pxs-def}).  The non-return probability $Q(x,t)$ is the
probability that a random walk, that starts at $x_0=1$, is at position
$x$ at time $t$, and that the origin is never visited.  In the
$t\to\infty$ limit, this quantity satisfies the diffusion equation
$Q_t=DQ_{xx}$ subject to the absorbing boundary condition
$Q(x=0,t)=0$.  This boundary condition ensures that only walks that do
not hit the origin are counted.  It is simple to construct $Q(x,t)$ by
the image method.  We merely place a random walk of opposite
``charge'' at $-x_0$; this construction ensures that the boundary
condition at $x=0$ is automatically satisfied.  In the continuum
limit, we have
\begin{eqnarray}
\label{F}
Q(x,t)&=& \frac{1}{\sqrt{4\pi Dt}}\,
\left[ e^{-(x-x_0)^2/4Dt}-e^{-(x+x_0)^2/4Dt}\right]
\nonumber \\
&\to & \frac{x}{\sqrt{4\pi D^3t^3}}\, e^{-x^2/4Dt},
\end{eqnarray}
with $x_0=1$.  The Laplace transform of this distribution is $Q(x,s)=
D^{-1}e^{-|x|\sqrt{s/D}}$ \cite{AS}.

In a similar vein, the $k^{\rm th}$-passage probability to the origin
at time $t$ is the convolution of a product of $k$ first-passage
probabilities at times \hbox{$t_1< t_2< \ldots <t_k$}.
Correspondingly, the Laplace transform for this $k^{\rm th}$-passage
probability is then the product of $k$ first-passage probabilities.
In turn, the first-passage probability to the origin is simply
\hbox{$F(0,s)= 1-[P(0,s)]^{-1}$}, where \hbox{$P(0,s)= 1/\sqrt{4sD}$}
is the Laplace transform of the occupation probability at the origin
\cite{W,fpp}.  This connection between the first-passage and
occupation probabilities is perhaps the most fundamental result in
first-passage statistics.  Thus the Laplace transform of the $k^{\rm
th}$-passage probability to the origin has the compact form
\cite{WR,W,fpp}:
\begin{eqnarray}
\label{Fmz} F^{(k)}(x=0,s)\to e^{-k\sqrt{4Ds}}.
\end{eqnarray}

Substituting these results in Eq.~(\ref{pxs-def}), the Laplace
transform for the probability distribution of the greedy random walk
is
\begin{equation}
\label{pxs}
G(x,s)\sim \sum_{k}  e^{-k\sqrt{s}}\,\,
\frac{1}{k^\alpha} \, e^{-|x|\sqrt{s}/k^\alpha}.
\end{equation}
Here we drop the first term $Q(x,s)$ in Eq.~(\ref{pxs-def}) because it
gives a subdominant contribution to $G(x,s)$.  To simplify the
derivations that follow, the diffusion coefficient is set equal to one
and all numerical prefactors of order one are ignored.  We perform the
sum over $k$ by first taking the continuum limit and then using the
Laplace method.  The integrand has a maximum at $k^*\propto
|x|^{1/(1+\alpha)}$.  We then expand the exponent function $f(k)=
-\sqrt{s}(k+ |x|/k^\alpha)$ to second order about this maximum and
perform the resulting Gaussian integral.  This gives
\begin{eqnarray*}
\label{pxs-result} G(x,s)\propto s^{-1/4} |x|^{(1-2\alpha)/2(\alpha+1)}\,\,
\exp\left(-s^{1/2} |x|^{1/(1+\alpha)}\right).
\end{eqnarray*}

Finally, we invert the Laplace transform by the Laplace method to
obtain, after straightforward steps, the probability distribution as a
function of time
\begin{equation}
\label{pxt-final} G(x,t)\sim \frac{|x|^{\frac{1-\alpha}{1+\alpha}}}{t}
\exp\left(-|x|^{2/(1+\alpha)}t^{-1}\right).
\end{equation}
Using the definition (\ref{fz}), the scaling function underlying the
probability distribution function $G(x,t)$ is
\begin{equation}
\label{fz-final}
\Phi(z) \sim |z|^{\frac{1-\alpha}{1+\alpha}}\,\,
\exp\left(-|z|^{2/(1+\alpha)}\right).
\end{equation}
Notice that for the particular case of $\alpha=1$, the probability
distribution is a pure exponential decay in $z$.

There are several features of this probability distribution worth
emphasizing.  First, it is straightforward to compute moments of the
displacement from (\ref{pxs}).  Thus, for example, we have (with the
diffusion coefficient restored)
\begin{eqnarray}
\label{xt}
\langle x^2(t)\rangle= \frac{1}{2}\frac{\Gamma(1+2\alpha)}{\Gamma(2+\alpha)}
(4D)^{1-\alpha} t^{1+\alpha}.
\end{eqnarray}
Notice that in the case of $\alpha=1$, the displacement of the greedy walk is
independent of the diffusion coefficient.  Second, for any non-zero value of
$\alpha<1$, no matter how small, greedy walks are eventually repelled from
the origin.  Third, the limit \hbox{$\alpha\to 0$} is singular.  This is
reflected in the limiting behavior of Eq.~(\ref{fz-final}),
\hbox{$\Phi(z)\sim |z|$}, when $z\to 0$.  In contrast, the probability
distribution of an ordinary random walk is finite at the origin.  Finally,
note that the cumulative distribution $C(x,t)=\sum_{x'>x}G(x',t)$ has the
scaling form $C(x,t)\to \Phi_C(x\,t^{-\nu})$, with the simpler scaling
function
\begin{equation}
\label{cumulative}
\Phi_C(z)\sim\exp\left(-|z|^{2/(1+\alpha)}\right).
\end{equation}
There is no algebraic prefactor in the cumulative distribution, and the
scaling function is finite at the origin, as it must.

\subsection{Simulations}

To test our predictions for the typical displacement and the
probability distribution, we turn to simulations.  For concreteness we
examined only the case of $\alpha=1$, where the step length increases
linearly in the number of returns to the origin (see Eq.~(\ref{ell})).
Our data are all based on averaging over $10^8$ walks.  As a
preliminary, we verified that the root-mean-square (rms) displacement
$x_{\rm rms}$ grows linearly with time in accordance with
Eq.~(\ref{xt}).

\begin{figure}[gt]
  \vspace*{0.cm}
  \includegraphics*[width=0.475\textwidth]{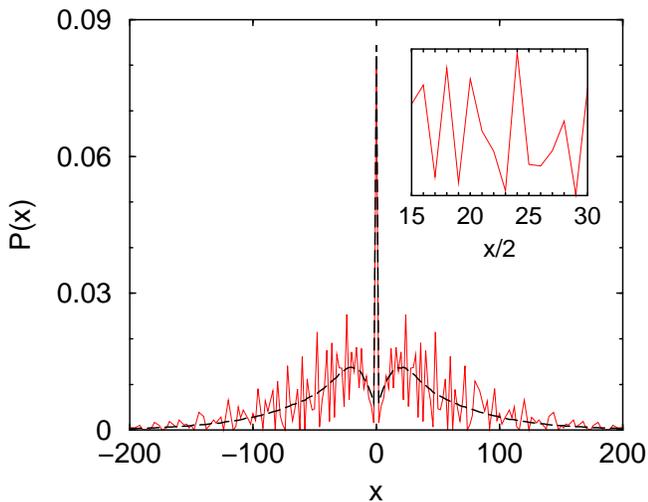}
\caption{Probability distribution of the greedy random walk after
  $t=10^2$ steps.  Shown are results for integer-valued step lengths (solid
  line) and continuum step lengths (dashed line).  The latter distribution
  was scaled up by a factor 2.  The inset shows detail of the prime-based
  fluctuations.}
 \label{px}
\end{figure}

The probability distribution itself exhibits several intriguing
features that lie outside of a scaling description (Fig.~\ref{px}).
First, there are huge fluctuations in the distribution.  These arise
because displacement values that are prime or have relatively few
prime factors are hard to reach by a greedy random walk.  While this
variability in the distribution is striking, it does not play a role
in the asymptotic scaling form of the distribution.

There is also a singularity at $x=0$ and secondary peaks at a distance
$t^{1/2}$ from the origin.  The singularity arises because the
probability of being exactly at the origin is not affected by the
enhancement mechanism of greedy walks.  All that is required is an
equal number of steps to the left and right, independent of when these
steps occur.  Thus the amplitude of this peak decays as $t^{-1/2}$, as
in a pure random walk.  In contrast, as follows from
(\ref{pxt-final}), the amplitude of the scaling part of the
distribution scales as $t^{-1}$.

To visualize the envelope of the distribution and the anomalous
behavior near the origin more clearly, we introduce a smoothed version
of the greedy random walk in which the step length grows by an
increment that is uniformly distributed between 0 and 2 upon each
return to the origin.  Such a construction still has a step length
that equals, on average, the number of returns to the origin, but
there are no longer any discreteness effects (Fig.~\ref{px}).  The
resulting probability distribution clearly reveals secondary peaks
close to the origin.  These are due to walks that never return to the
origin.  For this class of walks, the contribution to the probability
distribution in the continuum limit was given by Eq.~(\ref{F}).  Since
the characteristic length scale of this contribution is proportional
to $t^{1/2}$, the secondary peaks get squeezed toward the origin when
the distribution is plotted against the properly scaled coordinate
$x/t$.

\begin{figure}[t]
  \vspace*{0.cm}
  \includegraphics*[width=0.475\textwidth]{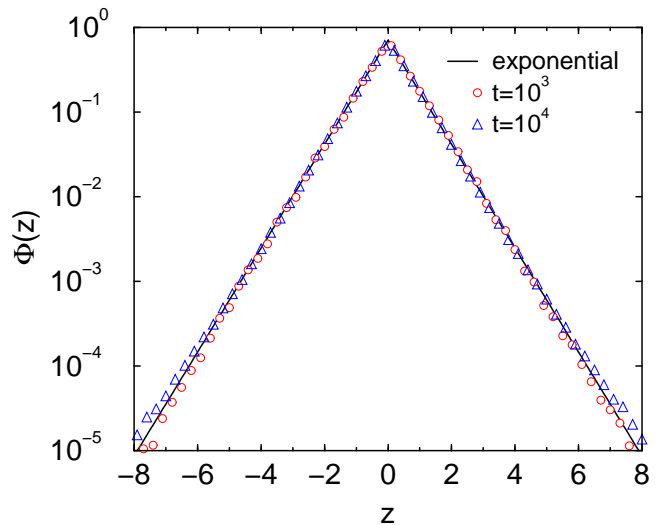}
 \caption{The scaling function of Eq.~(\ref{fz}) for walks with $10^3$ and
 $10^4$ steps on a semi-logarithmic scale.  Also shown as a reference is an
 exponential distribution.  The distributions are normalized to have a unit
 rms displacement, $\langle z^2\rangle =1$.}
 \label{fig-fz}
\end{figure}

More importantly, the continuum probability distribution of the greedy
walk obeys scaling, with the scaling function a purely exponential
function of the normalized displacement $z=x/x_{\rm rms}$, in
agreement with our analytic prediction given in Eq.~(\ref{fz-final}).

\section{Duration statistics}

We now turn to the question that motivated this work, namely, how long does a
game that is based on a greedy random walk last?  More abstractly, how long
does it take for a greedy random walk to escape the interval $[-L,L]$?  Our
basic result is that the typical escape time is relatively short, but higher
moments still involve the diffusive time scale $L^2$.

\begin{figure}[t]
  \vspace*{0.cm}
  \includegraphics*[width=0.475\textwidth]{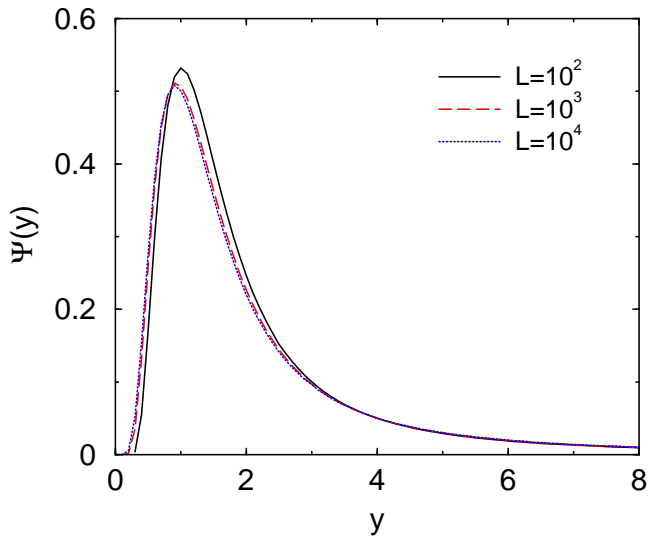}
 \caption{The scaling function of the exit-time distribution
 (Eq.~(\ref{psi-def})) from simulations with system size $L=10^2$ (solid
 line), $L=10^3$ (dashed line), and $L=10^4$ (dotted line).}
 \label{psiy}
\end{figure}

The displacement scaling law (\ref{typical}) suggests that the typical escape
time scales as
\begin{equation}
\label{typical-t}
t\sim L^{\mu}
\end{equation}
with $\mu=1/\nu=2/(\alpha+1)$.  The exit-time distribution, $E(t,L)$,
namely, the probability that the walk exits a system of size $L$ at
time $t$, should then follow scaling in the large-$L$ limit
(Fig.~\ref{psiy})
\begin{equation}
\label{psi-def}
E(t,L)\sim L^{-\mu}\,\Psi\left(t\,L^{-\mu}\right).
\end{equation}
Since $\mu<2$, the typical lifetime of a walk is shortened by the
greedy walk mechanism.

The minimal escape time is realized by hybrid zig-zag/ballistic walks
that have the maximal displacement.  Similar to the considerations
leading to Eq.~(\ref{xmax}), the escape time for such extremal walks
is $t=\tau+L(\tau/2)^{-\alpha}$, with $\tau$ the duration of the
zig-zag phase.  The escape time is minimized when $\tau\sim
L^{1/2\nu}\sim L^{\mu/2}$ and the minimal escape time also scales as $
L^{\mu/2}$.  Since the probability for such walks decays exponentially
with the escape time, we then infer from Eq.~(\ref{psi-def}), the
asymptotic behavior
\begin{equation}
\label{small-y}
\Psi(y)\sim \exp(-1/y),
\end{equation}
for $y\ll 1$.  This scaling is independent of $\alpha$ and thus, is
identical to that of an ordinary random walk.  We conclude that
short-lived games are relatively rare.

Long-lived walks are more interesting.  To determine the likelihood of
such walks, we need to understand greedy random walk trajectories
(Fig.~\ref{traj}) in finer detail.  First consider the return segments
of the greedy walk.  Upon the $j^{\rm th}$ return and after the next
step, the walk is at $\pm j^\alpha$.  Then from classic first-passage
properties \cite{F68,fpp}, the probability of returning again to the
origin without escaping is $(1-j^\alpha/L)$, while the probability of
escaping without another return is $j^\alpha/L$.  Notice that the
increase of the single-step length effectively reduces the system size
by the step length.  Using these results, the probability of returning
to the origin at least $k$ times is
\begin{eqnarray}
\label{prod}
R_k&=&\left(1-\frac{1}{L}\right)
\left(1-\frac{2^\alpha}{L}\right)\ldots
\left(1-\frac{k^\alpha}{L}\right)\nonumber\\
&\sim& \exp\left(-k^{1+\alpha}/L\right).
\end{eqnarray}
The typical number of returns to the origin before escape occurs is
found from the criterion $R_k\approx 1/2$; this gives
\begin{eqnarray*}
k\sim L^{1/(1+\alpha)}.
\end{eqnarray*}
This statement tell us the magnitude of the typical payoff in a greedy
random walk-based game.

\begin{figure}[t]
  \vspace*{0.cm}
  \includegraphics*[width=0.475\textwidth]{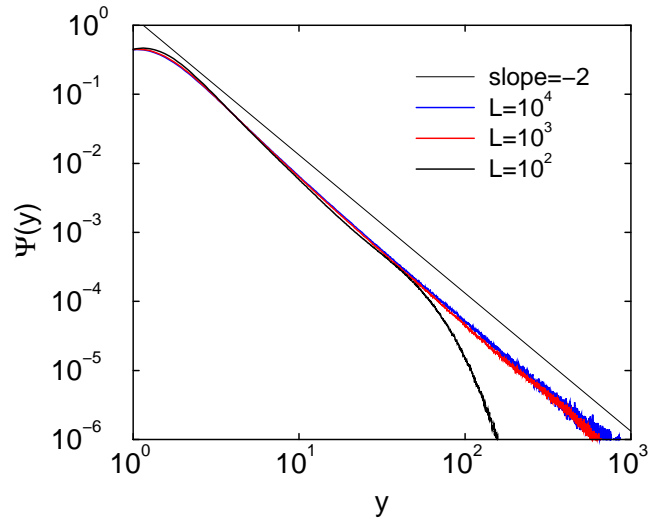}
 \caption{The scaling function of the exit-time distribution on a double
 logarithmic scale.}
 \label{psiy-tail}
\end{figure}

For a greedy walk to escape the system at time $t$, there may be
$k\geq 0$ returns to the origin, followed by a non-return segment.
From Eq.~(\ref{prod}), the probability for the former event is
$\exp\left(-k^{1+\alpha}/L\right)$.  In the long-time limit, the
probability for the non-return segment scales as
$(\frac{k^\alpha}{L})^3\, \exp\left(-tk^{2\alpha}/L^2\right)$, where,
as usual, we have ignored all factors of order one in the exponential.
The exponential term is just the controlling factor for the escape
probability of a random walk in an interval of length $L/k^\alpha$
\cite{fpp}.  The prefactor $(k^\alpha/L)^3$ ensures that the integral
of this factor over all time gives the correct ultimate escape
probability of $k^\alpha/L$.  Putting these elements together, the
probability $E(t,L)$ for a greedy walk to escape the interval $[-L,L]$
at time $t$ is
\begin{eqnarray}
\label{F-sum}
E(t,L)\sim \sum_{k=1} \left(\frac{k^\alpha}{L}\right)^3\, 
e^{-tk^{2\alpha}/L^2} \,\,
e^{-k^{1+\alpha}/L}.
\end{eqnarray}

By converting the sum to an integral, and noting that in the long time
limit, the last term is negligible, we obtain
\begin{eqnarray}
\label{large-t}
E(t,L)\sim \frac{L^{1/\alpha}}{t^{3/2+1/2\alpha}},
\end{eqnarray}
over the intermediate time range $L^{\mu}\ll t \ll L^2$.  Escape times
larger that $L^2$ are realized by walks that never return to the
origin.  However, such walks are exponentially unlikely,
$\exp(-t/L^2)$, so that their contribution is irrelevant in the
scaling limit.  In the limit $\alpha\to\infty$, walks may return to
the origin at most once, and the behavior $E(t,L)\sim t^{-3/2}$
follows from Eq.~(\ref{F}). Finally, the power-law form of $E(t,L)$
implies that the scaling function has the power-law decay for $y\gg 1$
(Fig.~\ref{psiy-tail})
\begin{eqnarray}
\Psi(y)\sim y^{-3/2-1/2\alpha}.
\end{eqnarray}

The existence of this power-law tail suggests that higher-order
moments of the distribution are not characterized by the typical
behavior (\ref{typical-t}).  By integrating Eq.~(\ref{large-t}) up to
the time scale $L^2$, we find three behaviors
\begin{eqnarray}
\langle t^n\rangle \sim 
\begin{cases}
L^{\mu\,n}&n<n_c;\\
L\ln L&n=n_c;\\
L^{2n-1}&n>n_c.
\end{cases}
\end{eqnarray}
Low-order moments are characterized by the typical escape time, while
high-order moments are described by pure diffusion.  At the boundary
$n_c=(\alpha+1)/2\alpha$, there is a logarithmic correction.  Thus
while the greedy walk mechanism significantly reduces the typical
duration of the game, there are substantial fluctuations in this
duration.  In the worst case, the duration would be proportional to
$L^2$, as in the ordinary random walk.

When the step length grows linearly with the number of returns to the
origin, Eq.~(\ref{large-t}) shows that the exit time distribution has
the asymptotic behavior $L/t^2$.  Thus the average lifetime of a
greedy walk includes the logarithmic correction
\begin{equation}
\langle t\rangle \propto L\ln L.  
\end{equation}
This behavior can be obtained directly by noticing that: (i) there are
of the order $k\sim L^{1/2}$ returns, (ii) the average $k^{\rm th}$
return time is $\langle t_k\rangle \propto L+L/2+\cdots L/k\propto
L\ln k$.

For completeness, we mention that the full survival probability can be
computed via the Laplace transform.  For an ordinary random walk
starting at $x=1$ in a domain of length $L$ with absorbing boundary
conditions, the Laplace transform of the first-passage probabilities
at $x=0$ and $x=L$ are, respectively \cite{fpp},
\begin{eqnarray*}
j_-(s)&=&\sinh [\sqrt{s/D}(L-1)]/\sinh [\sqrt{s/D}L]\\
j_+(s)&=&\sinh [\sqrt{s/D}]/\sinh[\sqrt{s/D}L].
\end{eqnarray*}
The first-passage probability at $x=L$ for the greedy random walk is
obtained by summing over the possible number of returns and scaling
down the domain size by the step length $l_k$ at each return.  This
gives
\begin{equation}
\label{E-prod}
E(s,L)=
\sum_{k=0}^\infty 
\frac{\sinh \sqrt{s}}{\sinh[\sqrt{s}\frac{L}{l_k}]}
\prod_{j=1}^{k} 
\frac{\sinh\left[\sqrt{s}\left(\frac{L}{l_j}-1\right)\right]}
{\sinh[\sqrt{s}\frac{L}{l_j}]}.
\end{equation}
The small-time and large-time behaviors given above follow from the
large-$s$ and small-$s$ behaviors of this expression, respectively.
Additionally, the $k$-th return probability given in Eq.~(\ref{prod})
equals the $s\to 0$ limit of the product in Eq.~(\ref{E-prod}).

\section{Summary}

The greedy random walk, in which the step length increases
algebraically with the number of returns to the origin, exhibits a
variety of unusual features. The distribution of displacements is
non-Gaussian, despite the fact that each epoch is characterized by a
Gaussian distribution. Similarly, the distribution of escape times
decays as a power-law tail even though each segment has an ordinary
exponential decay.

The most extended walks follow a two-stage process of immediate
reversals followed by a ballistic trajectory.  There are also several
features of the probability distribution that are outside of scaling,
including the contributions of walks that never return to the origin,
walks that are exactly at the origin, and large prime-number-induced
fluctuations.

The time for a greedy random walk to escape a finite interval is much
smaller than that of ordinary random walks.  However, because the
distribution of escape times decays as a power law in the long time
limit, there are large fluctuations in the escape time and the longest
possible games have a similar length to those based on ordinary random
walks.

We conclude that the greedy random walk provides a strategy for
completing zero-sum games quickly.  Increasing the stakes when the
game is tied accelerates the path to richness or to ruin.

\acknowledgments

We thank our children for being willing participants in our
machinations to speed up the game of war.  We also acknowledge DOE
grant W-7405-ENG-36 and NSF grant DMR0227670 for support of this work.

\end{document}